\documentclass[%
 reprint,
 amsmath,amssymb,
 aps,
]{revtex4-1}
\usepackage{graphicx}
\usepackage{dcolumn}
\usepackage{bm}
\usepackage{graphicx}
\usepackage{dcolumn}
\usepackage{bm}
\usepackage{bigints}
\usepackage{color}
\usepackage{amsmath,amssymb,graphicx}
\usepackage{amssymb}
\usepackage{gensymb}
\usepackage{amsfonts}
\usepackage{float}
\usepackage{braket}
\usepackage{tabularx}
\usepackage{tabulary}
\usepackage{tabu}
\usepackage{booktabs}
\usepackage{textcomp}

\newcolumntype{C}{>{\centering\arraybackslash}X}

\begin{document}
\preprint{APS/123-QED}
\title{The Impact of Spatial Correlation in Fluctuations of the Refractive Index on Rogue Wave Generation Probability}

\author{Mostafa Peysokhan$^{1,2}$}%
\author{John Keeney$^{1,2}$}%
\author{Arash Mafi$^{1,2,*}$}%

\email{mafi@unm.edu}
\affiliation{$^1$Department of Physics and Astronomy, University of New Mexico, Albuquerque, New Mexico 87131, USA \\
             $^2$Center for High Technology Materials, University of New Mexico, Albuquerque, New Mexico 87106, USA} 

\begin{abstract}
The presence of refractive index fluctuations in an optical medium can result in the generation of optical rogue waves. Using numerical simulations and statistical analysis, we have shown that the probability of optical rogue waves increases in the presence of spatial correlations in the fluctuations of the refractive index. We have analyzed the impact of the magnitude and the spatial correlation length of these fluctuations on the probability of optical rogue wave generation.
\end{abstract}
\maketitle
\section{Introduction}
Oceanic freak or rogue waves (RW) are large, rare, and unpredictable waves that appear in relatively calm waters, and are 
claimed to result in maritime disasters involving ship disappearances~\cite{muller2005rogue}. The science of RWs in optics 
was initiated by observing extraordinarily high field amplitude peaks at certain wavelengths in the chaotic spectrum from the 
supercontinuum~\cite{solli2007optical}. It is now understood that RWs can appear in different circumstances in
both linear and nonlinear systems~\cite{arecchi2011granularity,akhmediev2010editorial,Leonetti,Safari}. 
In recent years, numerous theoretical and 
experimental results have been reported on the generation and study of RWs in different optical platforms. Examples in nonlinear
systems include supercontinuum generation in optical fibers~\cite{solli2007optical,lafargue2009direct, solli2010seeded,buccoliero2011midinfrared}, 
nonlinear waves in optical cavities~\cite{montina2009non, residori2012rogue}, and Raman fiber amplifiers~\cite{finot2010selection,hammani2012experimental}.
There have been only a few reports on studies of RWs in linear 
systems~\cite{akhmediev2010editorial,hohmann2010freak,ying2011linear,mattheakis2016extreme,peysokhan2017optical}.
In particular, Ref.~\cite{hohmann2010freak} conducted a microwave transport experiment in a quasi-two-dimensional resonator with randomly distributed 
conical scatterers and found hot spots corresponding to RWs.

In this Letter, we expand on the results of Ref.~\cite{hohmann2010freak} and study the impact of the {\em spatial correlation in the fluctuations 
of the material dielectric constant} on the probability of RW generation. Our analysis targets the optical domain and the results are notable because in practice 
refractive index fluctuations always occur with a finite correlation length. In the examples that we explore in this Letter, we show that with increasing the correlation 
length, the probability of RW generation increases. We apply our analysis to examine whether the increased likelihood of hot spots (RWs) 
may be linked to premature material failure when exposed to high-intensity laser radiation~\cite{rudolph2013laser}.
Our analysis is performed in the linear domain, similar to Ref.~\cite{hohmann2010freak}. We study the effect of the correlation length ($\mu$) 
and the dielectric constant fluctuation contrast (DCFC, $\Delta\epsilon$), which is defined as the difference between the maximum and the minimum 
of the dielectric constant, on the probability of occurrence of optical RWs. We show that both the correlation length and DCFC have significant 
impacts on the probability of RW generation.

\section{Simulations}
Simulations are performed with the finite-difference time-domain (FDTD) method~\cite{fdtd} using Meep, which is an open-source 
software package for electromagnetics simulation~\cite{meep}. Meep simulates fully vectorial Maxwell's equations:
\begin{align}
\frac{\partial\mathbf{B}}{\partial t} &= -\nabla\times\mathbf{E} - \mathbf{J}_B - \sigma_B \mathbf{B},
&&\mathbf{B} = \mu \mathbf{H},\\
\frac{\partial\mathbf{D}}{\partial t} &= \nabla\times\mathbf{H} - \mathbf{J} - \sigma_D \mathbf{D},
&& \mathbf{D} = \varepsilon \mathbf{E},
\end{align}
where $\mathbf{E}$ is the electric field, $\mathbf{D}$ is displacement field, $\mathbf{H}$ is the magnetic field, and $\mathbf{B}$ 
is the magnetic flux density. $\varepsilon$ is the dielectric constant, $\mu$ is the magnetic permeability, 
$\mathbf{J}$ ($\mathbf{J}_B$) is the current density of electric (magnetic) charge, and $\sigma_B$ ($\sigma_D$) terms correspond 
to (frequency-independent) magnetic (electric) conductivities.

The occurrence of RWs is due to random fluctuations in the material; therefore, it must be studied using a statistical approach. For each dielectric medium
identified with a particular value of DCFC and correlation length, we create an ensemble of ``statistically equivalent'' dielectric media~\cite{mafi2015transverse}. 
Each ensemble contains a sufficiently large number of elements required for obtaining proper statistics to estimate the probability of RW generation.
The dielectric medium is assumed to have no absorption and is invariant in the y-direction, and light is incident along the z-direction. 
Both the width of the medium in the transverse x-direction and its length in the longitudinal z-direction are each chosen to be $100\lambda$.  
The incident light is a monochromatic plane-wave of wavelength $\lambda$, and the polarization of the light is perpendicular to the x-z plane.
Perfectly matched layers of thickness $\lambda$ are placed all around the medium. The numerical cell size is set at $\lambda/15$, which is sufficiently small 
compared with the wavelength of light to ensure the accuracy of the computation. The background dielectric constant of the medium is assumed to be
$\epsilon=2$. 
The propagation of the wave is computed until the steady state is established. 
Depending on the value of DCFC and correlation length, 2,000 to 5,000 simulations are performed in each case.
Simulations are performed on 20 nodes of a supercomputer, where each node has 8 CPUs.

To construct the medium with a randomly fluctuating dielectric constant with a specific correlation length, we employ the following procedure:
we first create a random vector $X$ representing the dielectric constant at each point on the 2D computational grid in the x-z plane. The dielectric constant 
on the computational grid is sampled at $N\,=\,$100 points in each spatial coordinate and is represented by indices $i,j=1,\cdots ,N$. 
An element of $X$ is represented by $X_\alpha$, where $\alpha(i,j)=i+N\times j$; therefore, $X$ has $100^2=10,000$ elements. 
$X_\alpha$ consists of uncorrelated random variables, which are chosen uniformly in the range $[-\Delta\epsilon,\Delta\epsilon]$; therefore,
$E(X_\alpha)\,=0$ and $E(X_\alpha,X_\beta)\,=\,\Delta\epsilon^2/3$, where $\alpha, \beta \,=,1,\cdots\,N^2$. In matrix language, we have:
\begin{equation}
E(X)\,=0,\quad E(X,X^T)\,=\,(\Delta\epsilon^2/3)\,{\mathbf 1},
\end{equation}
where $E$ represents the expectation value. To transform $X$ to a correlated random vector, we start with the desired covariance matrix $Q$ identified
by the correlation length $\mu$. The elements of $Q$ are constructed according to:
\begin{equation}
Q_{\alpha,\beta}\,=\,\exp(-\tilde{d}(\alpha,\beta)/\mu^2),
\end{equation}
where $\tilde{d}(\alpha,\beta)$ is the square of the physical distance between points $\alpha$ and $\beta$ on the computational grid.
Therefore, the correlation between two points decreases exponentially as a function of their separation. We then use Cholesky decomposition to decompose 
the real symmetric positive-definite matrix $Q$ 
into a unique product of a lower triangular matrix $L$ with its transpose $L^T$:
\begin{equation}
Q=LL^T.
\label{QLLT}
\end{equation}
We then define another vector $Z=LX$ with the following property:
\begin{equation}
E(Z)\,=0,\quad E(Z,Z^T)\,=\,(\Delta\epsilon^2/3)\,Q,
\end{equation}
which represents the desired correlated random vector.

To simulate the propagation of light and RW generation in each element of an ensemble for a given value of  $\Delta\epsilon$ and $\mu$, 
we generate a $Z$ vector, representing the fluctuations of the dielectric constant in that element. Although the variance of the elements 
of $Z$ are bounded by $\Delta\epsilon^2/3$, the size of the elements of $Z$ become occasionally larger than $\Delta\epsilon$, resulting 
in unacceptably large dielectric constant fluctuations. In such cases, we remove this element from the ensemble and generated another $Z$,
ensuring that the size of each element remains smaller than or equal to $\Delta\epsilon$. By following this procedure, the
correlation is persevered and the maximum values of fluctuation in the dielectric constant is guaranteed to remain 
below $\Delta\epsilon$ for a more realistic representation of a dielectric. Because of this selective elimination, the statistics will 
deviate slightly from the continuous uniform distribution, which is inconsequential in the general observations we present in this Letter. 

\section{Results}
\begin{figure}[t]
 \centering
 \includegraphics[width=3.4in]{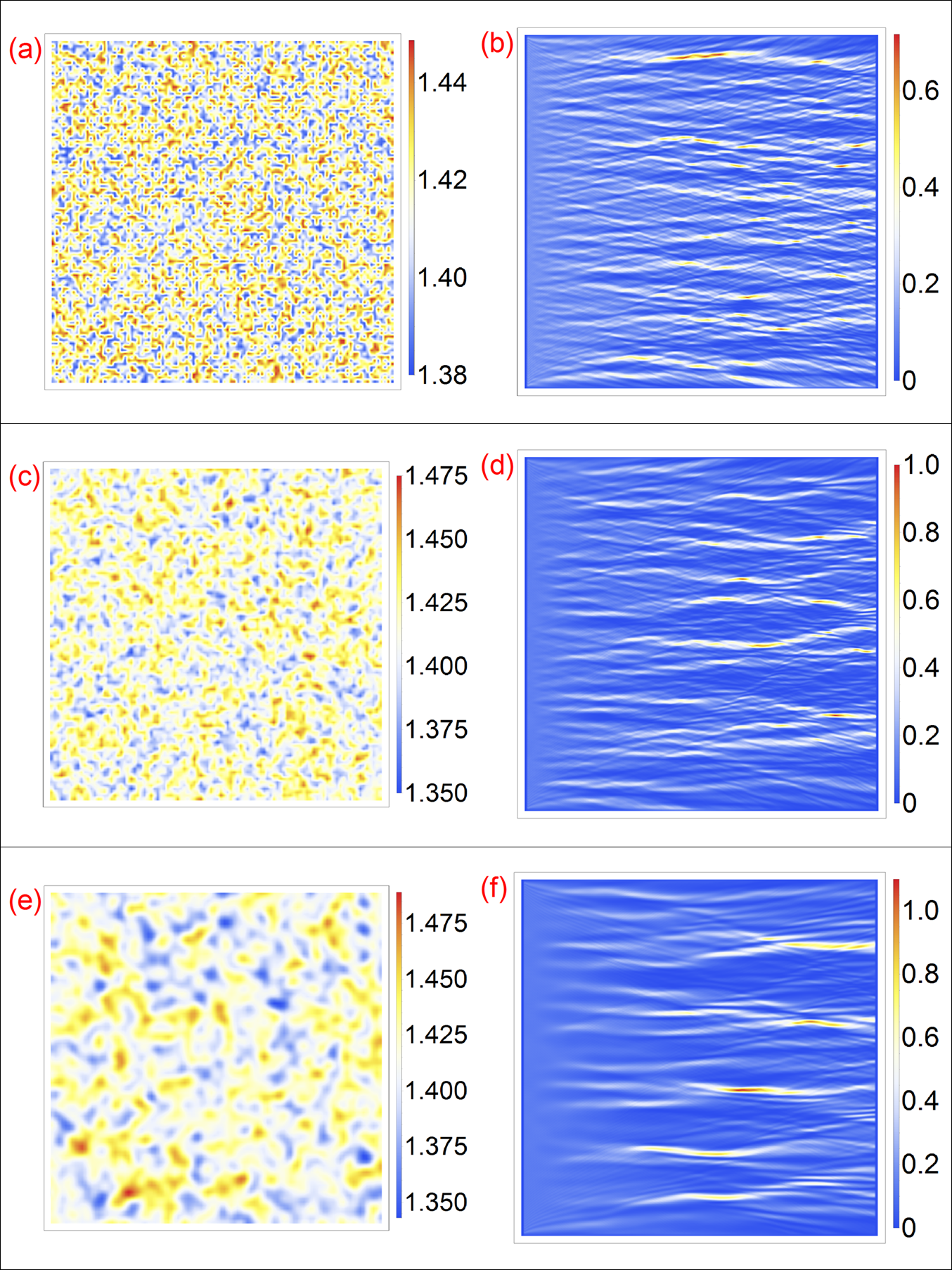}
 \caption{(a), (c), and (e) show the fluctuations in the refractive index profiles with the correlation 
lengths of $0.2 \lambda$, $1.3 \lambda$, and $2.8 \lambda$, respectively, where $\Delta\epsilon=0.4$ is assumed in all three cases.  
(b), (d), and (f) show the light intensities related to (a), (c), and (e), respectively.
The incident plane-wave propagates along the z-direction, from left to right. }
 \label{fig:correlation-propagation}
\end{figure}

In Fig.~\ref{fig:correlation-propagation}, the refractive index and light intensity profiles are shown for three different dielectric media
with different values of the correlation length for refractive index fluctuations. From the top row to the bottom row, the correlation lengths
are $0.2\lambda$, $1.3\lambda$, and $2.8\lambda$, respectively; $\Delta\epsilon=0.4$ is assumed in all three cases. The left
column in Fig.~\ref{fig:correlation-propagation} shows the refractive index fluctuations and the effect of the increased correlation length
from top to bottom row is visible. The right column shows the intensity of the propagated light: in each case, the incident light 
is a plane-wave with the wavelength $\lambda$, which propagates along the z-direction (from left to right). The branching flow of light 
clearly appears in this linear regime, and several high-intensity spots are observed in each case. 
As the correlation length is increased, the number of branches decrease, but the peaks appear to become more prominent and intense.
Therefore, it is desirable to investigate the impact of an increase in the correlation length of the refractive index fluctuation
on the occurrence of the RWs quantitatively. 

\begin{figure}[htp]
 \centering
 \includegraphics[width=3.4in]{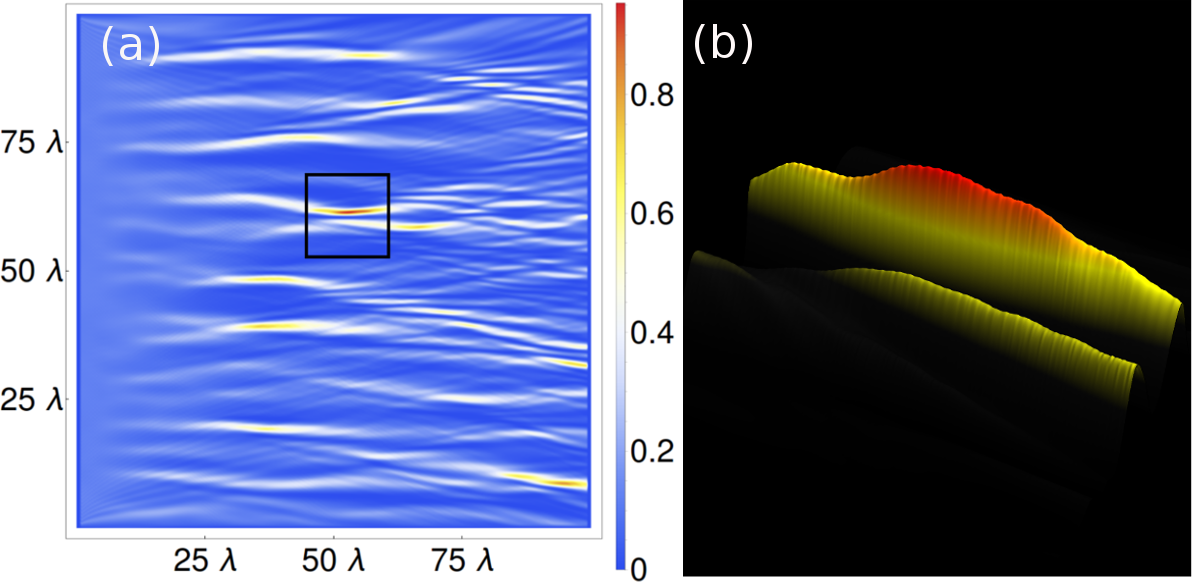}
 \caption{(a) The propagation of light through a random dielectric medium with $\mu=2.8 \lambda$ and $\Delta\epsilon=0.4$ is shown. 
          The incident light is a plane-wave with the wavelength $\lambda$. The light is propagated along the z-direction (from left to right). 
          (b) A 3D display of the intensity corresponding to the black square region in subfigure (a) displays the profile of an optical RW.}
 \label{fig:sample-simulation-1}
\end{figure}

To quantify the occurrence rate of the RWs, we first need to identify them. In Fig.~\ref{fig:sample-simulation-1}, we show an example of the 
propagation of light through a randomly correlated dielectric medium, defined by $\Delta\epsilon=0.4$ and $\mu=2.8 \lambda$.
Figure~\ref{fig:sample-simulation-1}(b) is a three-dimensional display of intensity, zoomed-in over the black square 
in Fig.~\ref{fig:sample-simulation-1}(a), in which the optical RW is seen.
To find the occurrence of an RW event, after steady-state of the propagated wave is established, the intensity of the optical field is extracted
in the matrix format sampled at each position. Then a two-dimensional local intensity peak-detection procedure is performed over the intensity matrix,
where each intensity pixel is compared to its eight neighbors, from which a peak intensity probability distribution (PIPD) is calculated.
The PIPD, $P(I)$, is defined such that $P(I)\,dI$ is proportional to the number of local peaks in the intensity range of $I$ and $I+dI$.

\begin{figure}[t]
\centering
\includegraphics[width=3.3in]{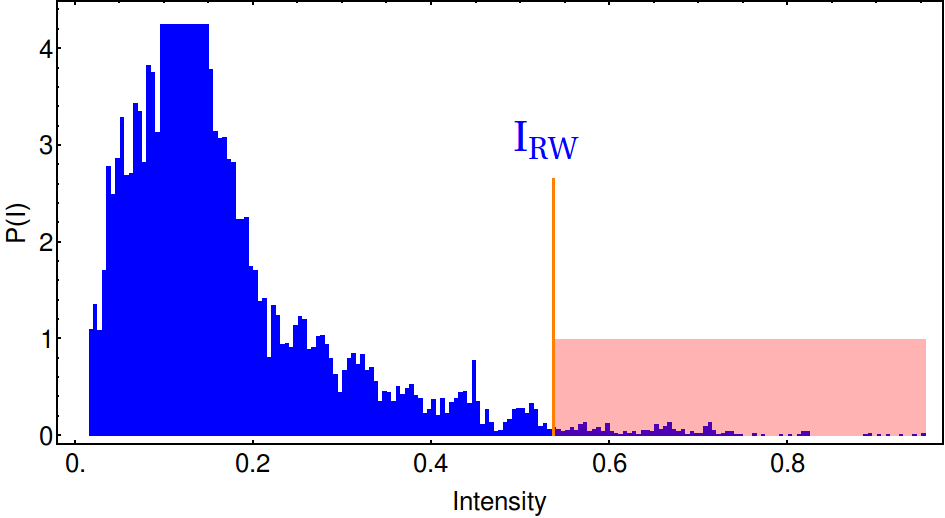}
\caption{The simulations presented in this figure correspond to $\mu=2.8 \lambda$ and $\Delta\epsilon=0.4$.
         The probability distribution of the peak intensities is plotted versus the intensity values of the peaks. 
         The vertical orange line marked with $I_{RW}$ indicates the threshold of the optical RWs based on {\em Method 2}, 
         above which the events are identified as RWs.}
\label{fig:Hist}
\end{figure}

To identify an intensity peak as a RW after PIPD is determined, one can use one of the following two commonly used methods:
{\em Method 1} is based on the Longuet-Higgins random seas model~\cite{Longuet-Higgins}, where the central limit theorem is 
applied to the random superposition of a large number of optical waves with different propagating directions, resulting in a 
Rayleigh probability distribution for the intensity peaks of the form $P(I)= e^{-I}$; and in {\em Method 2}  
RWs are identified using the RW-intensity-threshold in PIPD. The RW-intensity-threshold ($I_{\rm RW}$) is given
by  $I_{\rm RW} = 2 I_S$, where $I_S$ is the mean of the upper third of peaks in the PIPD
and all peak events with higher intensity than $I_{\rm RW}$ are identified as RWs~\cite{dudley2014instabilities}.
In Fig.~\ref{fig:Hist}, we employ {\em Method 2} to the configuration discussed in Fig.~\ref{fig:sample-simulation-1}:
the calculated PIPD is shown, where the peak events that satisfy $I_{\rm peak}> I_{\rm RW}$ are highlighted by a darker background color.

\begin{figure}[t]
 \centering
 \includegraphics[width=8 cm]{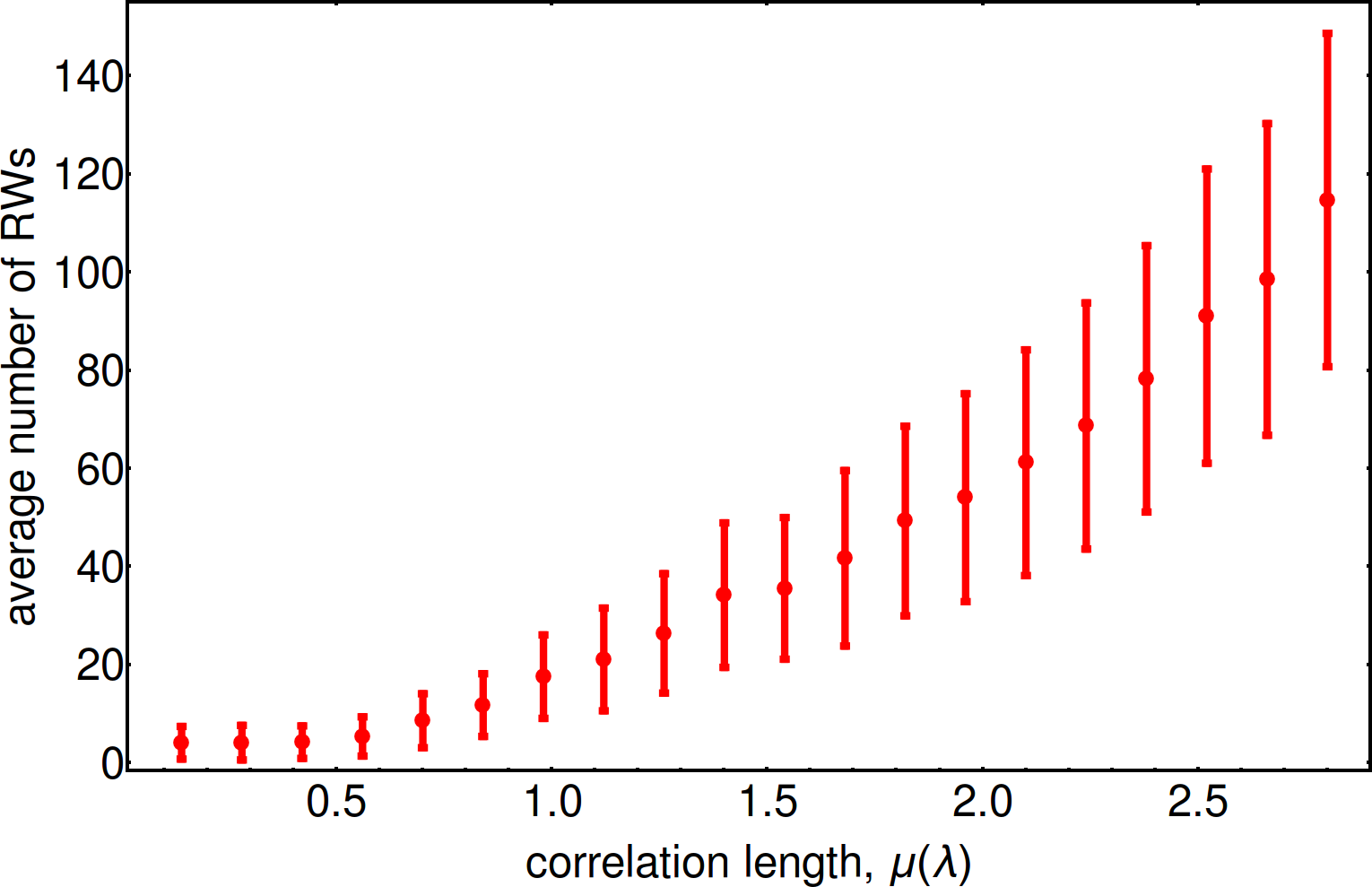}
 \caption{The number of RWs versus the correlation length: the dots are the average number of RWs for a specific correlation length and the error 
bars are the standard deviation of the total number of RWs for each correlation length.}
 \label{fig2}
\end{figure}

In this Letter, we use {\em Method 2}, which is also commonly used in the theory of scattering from random potentials.
To evaluate the impact of the correlation length on the occurrence probability of RWs, we propagated an optical 
plane wave through a randomly correlated material with the same geometry and the dielectric contrast of $\Delta\epsilon=0.08$.
The number of RWs is counted based on the statistical approach of {\em Method 2}. This process is repeated for 2000 different 
media (random realizations of the ensemble) for each value of the correlation length. Finally, the average and standard deviation 
of the number of RWs are calculated and the result is shown in Fig.~\ref{fig2}. The average number of RWs increases by
nearly two orders of magnitude when the correlation length increases from $0.2 \lambda$ to $2.8 \lambda$--this implies that 
the effect of correlation length is substantial on the probability of RW generation. This observation is the main finding in 
this Letter and the simulations are generic enough to suggest that this behavior is universal for at least a broad range of 
correlation lengths.

\begin{figure}[t]
 \centering
 \includegraphics[width=8 cm]{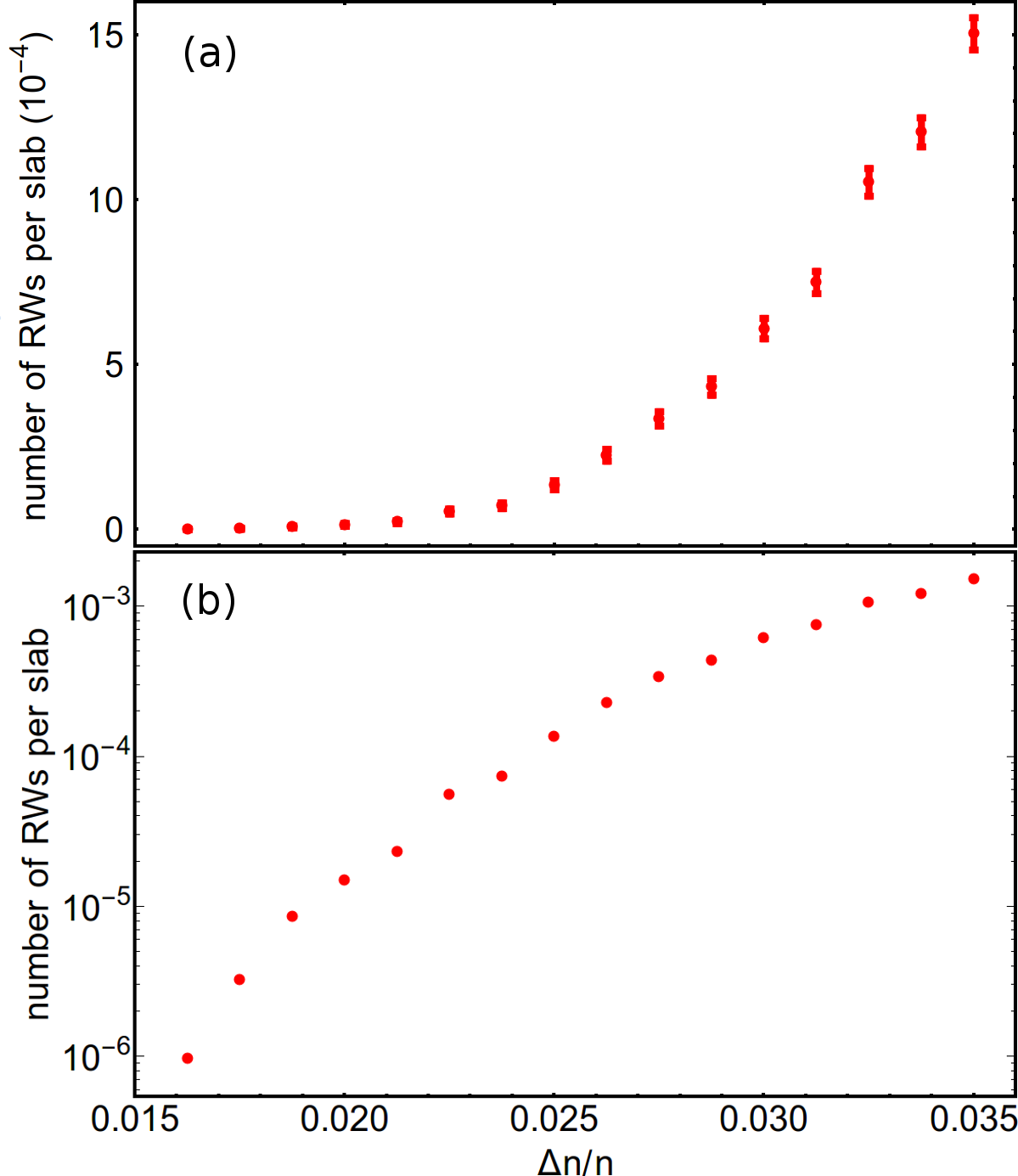}
 \caption{(a) The mean number of RWs per simulated slab as a function of the normalized refractive index fluctuation contrast $\Delta n/n$ is shown, where 
          the correlation length of $\mu=2.8 \lambda$ is assumed. The error bars are the standard error calculated for a sample size of 5,000 at each point;
          (b) Identical to (a), except plotted in a logarithmic scale.}
 \label{Fig:n-range}
\end{figure}

Another parameter that plays an essential role in forming optical RWs is the fluctuation contrast in the dielectric constant. To investigate its impact, 
we propagated an optical plane wave through a randomly correlated material with the same geometry and the
correlation length of $\mu=2.8 \lambda$ for different values of $\Delta\epsilon$. For each value of $\Delta\epsilon$, a large number of
ensemble elements were generated to obtain sufficient statistics on the number of RWs. The number of ensemble elements was 5,000 for 
each value of $\Delta\epsilon\approx 2n\Delta n$, where $\Delta n$ is the fluctuation contrast in the refractive index. 
The scaling of the number of RWs generated versus $\Delta n/n$ is shown in Fig.~\ref{Fig:n-range}(a) in a linear scale and in 
Fig.~\ref{Fig:n-range}(b) in logarithmic scale and the results indicate that the effect of the magnitude of fluctuation in the refractive index has
a sizable impact on the probability of RW generation.  

\begin{figure}[htp]
 \centering
 \includegraphics[width=8 cm]{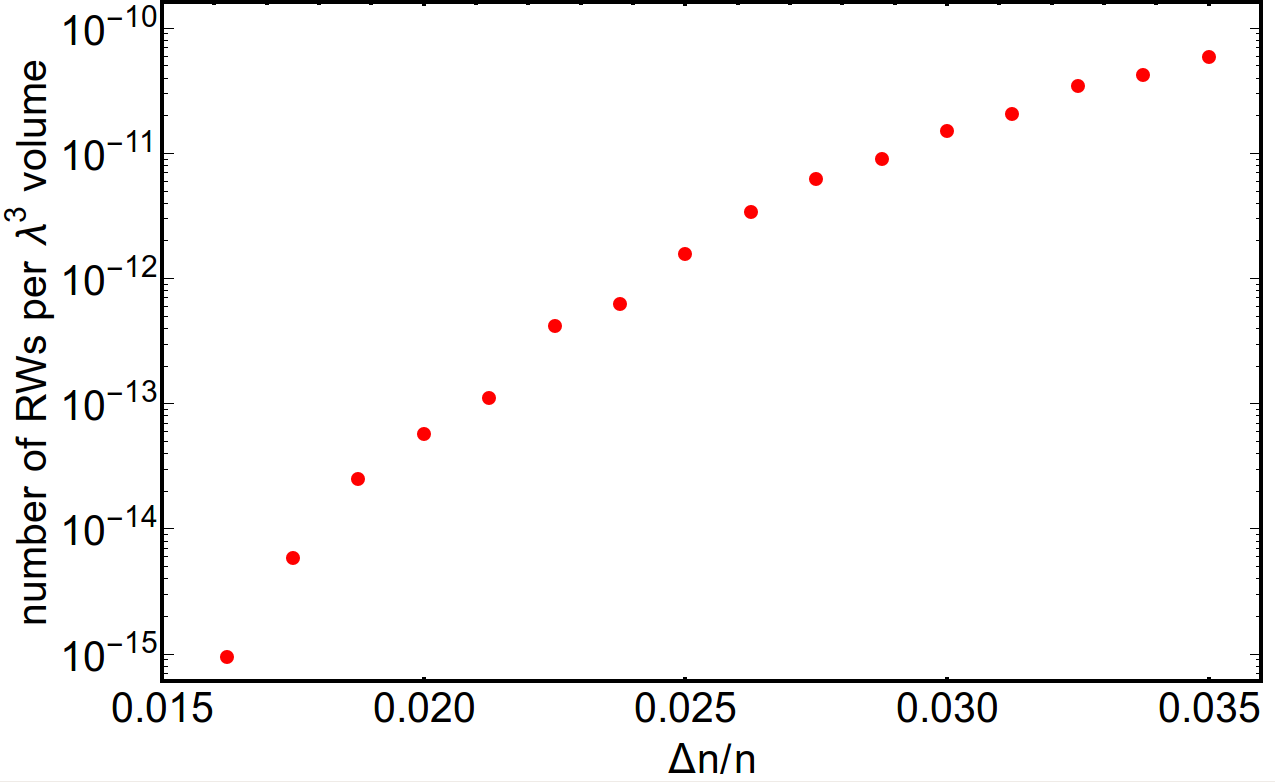}
 \caption{The mean number of RWs per $\lambda^3$ volume as a function of the normalized refractive index fluctuation contrast $\Delta n/n$ is shown, where 
          the correlation length of $\mu=2.8 \lambda$ is assumed.}
 \label{Fig:n-range-3}
\end{figure}

The results presented in Fig.~\ref{Fig:n-range} are for in-plane fluctuations. In reality, the refractive
index fluctuations occur in the full volume; therefore, one expects that the number of RWs in a sample with fluctuations 
in three dimensions ($N_3$) scale as $N_3\propto N_2^{3/2}$, where $N_2$ is the number of RWs in a sample with only in-plane fluctuations.
Considering that the area of each sample here is assumed to be $10^4\lambda^2$, one can readily estimate the mean number of RWs generated
per $\lambda^3$ volume element in the presence of the correlated refractive index fluctuations in 3D. The result is plotted in
Fig.~\ref{Fig:n-range-3}, which is extracted from Fig.~\ref{Fig:n-range}(b), where the number of RWs is divided by $10^4$ to find 
the number of in-plane RWs per $\lambda^2$ area and subsequently raised to the power $3/2$ to find the expected number of RWs
per $\lambda^3$ volume.

Figure~\ref{Fig:n-range-3} shows that for $\Delta n\approx 0.02$, the mean number of RWs per $\lambda^3$ volume is on the order of $10^{-15}$. 
Therefore, for visible wavelength with $\lambda\approx 500$\,nm, the mean number of RWs generated in a sample of 1\,cm$^3$ volume is approximately
0.01. In optical glasses, the nominal value for refractive index fluctuations is around $\Delta n\approx 0.00001$ or less. Figure~\ref{Fig:n-range-3} 
indicates that the probability for RW generation at such small refractive index fluctuations is quite small. Therefore, RWs are likely not
responsible for the premature failure of optical materials below their nominal optical damage threshold in the bulk. However, it is possible that
the presence of contaminants on optical surfaces such as dust, water, and skin oils, or imperfect polishing provide sufficient phase fluctuation
on the surface to create hot spots as RWs near the surface, resulting in a lower damage threshold. It must be noted that the discussions in this Letter
are exclusively on RW generation according to the definition of {\em Method 2} and it is quite possible that refractive index fluctuations of 
around $\Delta n\approx 0.00001$ can still result in intensity peaks that do not qualify as a RW but can still contribute to the premature 
optical material failure.
\section{Conclusion}
We have shown that the presence of spatial correlation in the fluctuations of the refractive index of a dielectric media has an impact on the probability of rogue wave generation. In the examples that we explored in this Letter, we observed that by increasing the correlation length, the likelihood of rogue wave generation increases. A possible explanation for this behavior is that the presence of a spatial correlation result in an underlying structure that mimics a photonic crystal, in which the rogue waves are created as a coherent enhancement of amplitude due to a Bragg-type scattering behavior~\cite{Chutinan:99,Mafi:15}. Our examples are generic; therefore, we believe our conclusions can be broadly applied to all coherent wave systems.   
\section*{Acknowledgments}
The authors would like to acknowledge partial support from the Office of the Vice President for Research – University of New Mexico (UNM). 
A.M. acknowledges partial support from National Science Foundation (NSF) (1807857,1522933) and J.K.  acknowledges partial support from
NSF (1522933). Authors are grateful to Prof. Kevin Malloy at UNM for illuminating discussions, Dr. Ryan Johnson and the UNM Center 
for Advanced Research Computing for large-scale computational resources used in this work, and Dr. Ardavan Oskooi for help with the MEEP software.

\end{document}